\documentclass[prl,preprint,preprintnumbers,amsmath,amssymb,floats]{revtex4}
\usepackage{graphicx}
\usepackage{dcolumn}
\usepackage{bm,epstopdf,soul}
\renewcommand{\abstractname}{}    

\usepackage{amsmath}
\usepackage{color}
\usepackage{soul}

\usepackage{times}

\usepackage{amsmath}
\usepackage{color}
\usepackage{soul}

\def\baselinestretch{1.5}

\begin{document}


\title{Observation of high-temperature quantum anomalous Hall regime in intrinsic MnBi$_2$Te$_4$/Bi$_2$Te$_3$ superlattice}



\author{Haiming Deng$^1$, Zhiyi Chen$^1$, Agnieszka Wo{\l}o\'{s}$^2$, Marcin Konczykowski$^3$, Kamil Sobczak$^{2,4}$, Joanna Sitnicka$^2$, Irina V. Fedorchenko$^5$, Jolanta Borysiuk$^2$, Tristan Heider$^6$, {\L}ukasz Pluci\'{n}ski$^6$, Kyungwha Park$^7$, Alexandru B. Georgescu$^8$, Jennifer Cano$^{8,9}$ \& Lia Krusin-Elbaum$^1$}

\vspace{0.5\baselineskip}
\def\baselinestretch{1.2}
\affiliation{$^1$Department of Physics, The City College of New York - CUNY, New York 10031, USA}
\affiliation{$^2$Faculty of Physics, University of Warsaw, 00-681 Warsaw, Poland}
\affiliation{$^3$Laboratoire des Solides Irradi\'{e}s, \'{E}cole Polytechnique, CNRS, CEA, Universit\'{e} Paris-Saclay, 91128 Palaiseau cedex, France}
\affiliation{$^4$Faculty of Chemistry, University of Warsaw, 02-089 Warsaw, Poland}
\affiliation{$^5$Kurnakov Institute of General and Inorganic Chemistry, Russian Academy of Sciences, 119991 Moscow, Russia}
\affiliation{$^6$Forschungszentrum J\"{u}lich GmbH, 52425 J\"{u}lich, Germany}
\affiliation{$^7$Department of Physics, Virginia Tech, Blacksburg, Virginia 24061, USA}
\affiliation{$^8$Center for Computational Quantum Physics, Flatiron Institute, New York, New York 10010, USA}
\affiliation{$^9$Department of Physics and Astronomy, Stony Brook University, Stony Brook, New York 11794, USA}


\begin{abstract}
	\def\baselinestretch{1.2}
	\vspace{1mm}
	\noindent \small \textbf{The quantum anomalous Hall effect \cite{QAHReview-SCZhang2016,QAH-realMat-2016} is a fundamental transport response of a topologically nontrivial system in zero magnetic field.
		Its physical origin relies on the intrinsically inverted electronic band structure and ferromagnetism \cite{Yu-QAHtheory2010}, and its most consequential manifestation is the dissipation-free {flow of chiral charge currents} at the edges \cite{AHE-Nagaosa2010} that can potentially transform future quantum electronics \cite{chSC-Qi-Hughes-Zhang2010,AFMspintronics2018}.
		Here we report a previously unknown Berry-curvature-driven \cite{AHE-Nagaosa2010,BerryPhaseReview2010} anomalous Hall regime {(`Q-window')} at above-Kelvin temperatures in the magnetic topological bulk  crystals where through growth Mn ions self-organize into a period-ordered MnBi$_2$Te$_4$/Bi$_2$Te$_3$ superlattice. Robust ferromagnetism of the MnBi$_2$Te$_4$ monolayers
		opens a large surface gap \cite{Rienks2019,Otrokov2019}, and anomalous Hall conductance reaches an $e^2/h$ quantization plateau when the Fermi level is tuned into this gap within a Q-window in which the anomalous Hall conductance from the bulk is to a high precision zero.
		The quantization in this new regime is not obstructed by the bulk conduction channels and thus should be present  in a broad family of topological magnets.
	}
\end{abstract}




\maketitle
\normalsize



In the quantum Hall (QH) effect --- a quantized version of the conventional (normal) Hall effect  ---  the Hall conductance, arising from a transverse voltage generated by the longitudinal current, has a profound topological origin \cite{Thouless1982}. Consequently, it takes on quantized values dependent only on fundamental constants: $h$ (Planck's constant) and $e$ (electron charge).
When a two-dimensional (2D) electron system is under high magnetic field, electron orbits are quantized into Landau levels, engendering an insulating state with dissipationless chiral currents flowing around the edges.
Distinct from QH is the quantized anomalous Hall (QAH) effect \cite{QAHReview-SCZhang2016}, which occurs in zero field without Landau quantization. Instead, it requires breaking time-reversal symmetry and a topologically nontrivial bandstructure with the Fermi level inside the bandgap. In a
2D QAH insulator, the Hall conductance is $G_{yx} = C e^2/h$, where $C$ is an integer topological invariant called the Chern number \cite{Thouless1982,AHE-Nagaosa2010} originating from the Berry curvature \cite{BerryPhaseReview2010} in momentum space.

{Among diverse systems proposed to host QAH \cite{Haldane1988,QAH-HgCr2Se4-2011,QAH-realMat-2016,Yu-QAHtheory2010}
	magnetic topological insulators \cite{Yu-QAHtheory2010}, formed by an intricate interplay between spin-orbit interactions \cite{Qi2011} and induced intrinsic magnetism, appeared most promising. Indeed, the first experimental realization of QAH was
	in Cr-doped ($x = 0.15$) ultrathin epitaxial films} of (Bi,Sb)$_{2-x}$Te$_3$ \cite{CZC-QAH1st_exp2013}.
The out-of-plane ferromagnetism induced by doping with
Cr (or V) breaks TRS and opens a small ($5-10~\textrm{meV}$) Dirac mass gap \cite{ChenMassiveDirac2010} in the topological surface states and the {Fermi level is then fine-tuned into this gap by electrostatic gating.} Quantization was observed in a limited ($\sim 5-10~ \textrm{nm}$) thickness range \cite{Feng-thickness2016},
with the lower limit fundamentally set by hybridization of the top and bottom surface states.
{The upper limit depends on the top and bottom `asymmetry', and
	on the nature of magnetic order  \cite{WeidaWu2018}.
	It is widely acknowledged that the ubiquitous magnetic doping inhomogeneities and Bi-Sb alloying disorders
	\cite{JCDavis2015} set the magnetization $M(T)$ in such films to be non-mean-field-like, with the onset of anomalous
	Hall conductance $G_{yx}(T)$ in the low sub-Kelvin range \cite{QAH-scZhang-review2016,Kou2015}, downward-lagging the ferromagnetic Curie temperature $T_C \sim 30~\textrm{K}$} by orders of magnitude.
Thus far, QAH has been realized in only one material system: (Cr,V)-doped BST thin films at these extremely low temperatures.

Here we report a discovery of a much higher-temperature QAH {regime} in a low-disorder bulk topological material, {where under `dilute' magnetic element concentration conditions, the crystalline order and the magnetic structure {(Figs.~1a-c, and {S1})} are self-organized into a well-ordered topological superlattice. {The quantization is observed when the Fermi level aligns with the Dirac mass gap.
		This occurs within an energy window in the conduction band (`Q-window') in which bulk contribution to the anomalous Hall conductance is null --- {a regime not previously considered or accessed in thin films.}}
	
	The system is nominally a canonical topological insulator (TI) Bi$_2$Te$_3$ with a small ($2\textrm{at}\%$ in total) population of Mn ions, grown as $n$-type (electron conductor) \cite{Scanlon-antisites2012}, see Methods.
	The magnetization aligns out-of-plane and the ferromagnetic $T_C$ does not depend on carrier density or type.}
The well-known bandstructure of undoped Bi$_2$Te$_3$ \cite{Zhang-NatPhys09} is not an {obvious}
candidate for QAH, since the Dirac point, nestled in the bulk valence bands,
would be largely unaffected by a random distribution of Mn in the dilute limit \cite{Hor-ptype_MnBT-2010}.
{However, when the growth process is properly designed (see Methods)}, Mn ions do not distribute uniformly. Instead, the quintuple-layered (QL) crystal structure
of undoped Bi$_2$Te$_3$ \cite{Chen2009} is modified from a disordered magnetic impurity system into a nearly-periodic
alternating sequence of QLs (Bi$_2$Te$_3$) and septuple layers (SL) of atoms, each SL with the MnBi$_2$Te$_4$ crystal structure \cite{MBT-structure2013} {(Fig.~1c).}
{$\textrm{Mn}^{2+}$ ions are predominantly incorporated into a monolayer within each SL (Figs. 1a,b), with only sparse amounts populating QLs and with  SLs roughly separated by four QLs (Figs.~1a and {S1}).} A strong exchange coupling of the Mn ions within SLs establishes {out-of-plane ferromagnetism} with $T_C \cong 13~K$ (Figs.~1e, {S2a, and S2b}),
with $\textrm{Mn}^{2+}$ within QLs spin-polarized by the intimate proximity to SLs ({Fig.~S3}).
{The magnetization $M(T)$ of our crystals sharply rises just below $T_C$ {(Fig.~1e)}, in a remarkably close correspondence with the temperature dependence of anomalous Hall conductance $G_{yx}^{AHE}(T)$ {(Fig.~1e)}, which also onsets at $T_C$.} The approach to quantized value $e^2/h$ on cooling, shown here for a thick ($300~\textrm{nm}$) crystal (sample S1), is evident at temperatures of several Kelvin. A nearly-square
zero-field quantized $G_{yx}^{AHE}(H)$ hysteresis loop
and {the corresponding change in the longitudinal magnetoconductance $G_{xx}^{AHE}(H)$ are shown in {Figs.~1f, g.}}


{The electronic bandstructure of our self-organized Mn-superlattice in Bi$_2$Te$_3$ is very different from that of dilute randomly-doped or undoped Bi$_2$Te$_3$. Recent density functional theory (DFT) {calculations \cite{Rienks2019,Otrokov2017} of the surface bands in thin MnBi$_2$Te$_4$/Bi$_2$Te$_3$ (SL/QL) heterostructures and bilayers, without accounting for the ever present vacancies}, predict a much enlarged {($\gtrsim 70~\textrm{meV}$)} Dirac gap located roughly in the middle of the bulk gap. {A large Dirac gap is also seen in a pure SL material, which is antiferromagnetic, by the angularly resolved photoemission spectroscopy (ARPES) \cite{ARPES-SL-NC2019,Otrokov2019}.} {This large gap is lacking in dilute randomly doped Bi$_2$Te$_3$.
		
		In order to reach the anomalous quantum Hall regime, the Fermi level should be positioned within the surface Dirac gap. However, the as-grown crystals are initially strongly \textit{n}-type.  Therefore,} to tune the surface bandstructure to quantization we utilize a two-step vacancy engineering process {(see Figs. S4 and S5)} --- a technique we have demonstrated previously \cite{irrad-Lukas2016} in a variety of TIs. First, irradiation with high-energy ($\sim 2.5$ MeV energy) electrons is used to create a uniform distribution of (predominantly donor) vacancies \cite{Scanlon-antisites2012}, making the material less \textit{n}-type (Fig. S5a). Next, thermal annealing is used to shift the surface bands upward relative to the bulk bands. By annealing at a series of temperatures and measuring the low-temperature longitudinal and Hall conductance between each step, we are able to fine-tune the process and track the evolution of the bandstructure. The vacancy-driven bandstructure modification we employ here to tune the system to quantization is supported by the DFT calculations that incorporate vacancies {{{(Fig. {S6}}}} and Supplementary Section \textbf{B}) and is consistent with ARPES {(Fig. {S7}).}
	We also note that in
	\textit{randomly} Mn-doped (2at\%) Bi$_2$Te$_3$ crystals without the intrinsic superlattice (see Methods), the irradiation induces the expected $p$-to-$n$ conversion ({Figs.~S4a, b}).
	Notably, the anomalous Hall effect in this case is null {on the \textit{p}-type side  ({Fig.~S4c})} and small all through the conversion (see also Ref. \cite{Checkelsky-Mndoped-carrier-dep2012}), even though the magnetization is still carrier independent and large ({Fig.~S4d}). This observation brings into focus the critical role played by the bandstructure  and signals the importance of Berry curvature effects.}

{Figures 2a and 2b schematically depict the thermal annealing process. During annealing, vacancies diffuse from the bulk to the surfaces, where they are expelled, reducing the bulk doping and moving the Fermi level downward. The surface defect 'evaporation' \cite{Vac-mag2019} also establishes a vacancy gradient (Fig. 2a), causing surface band bending which results in formation of subsurface 2D electron gas (2DEG) quantum-well-like \cite{YYang-thickness2010} states due to quantum confinement \cite{Bahramy2012}.
	Figure 2b depicts these bands at four steps during the annealing process. Initially (step 1) the Dirac bands (blue lines) and 2DEG bands (grey lines) are aligned with the bulk bands (grey shaded areas), and the Fermi level is within the bulk conduction band (BCB) at position $E_F^0$.  Upon annealing, the Fermi level shifts slowly downward, while at the same time the surface bands rapidly shift upward relative to the bulk bands. Thus, we can expect that with annealing, the Fermi level will first pass into the 2DEG separation $E_{2D}$ (point 2), then into the surface Dirac gap $\Delta$ (point 3), and finally into the 2DEG and Dirac valence bands (point 4), all while the bulk remains \textit{n}-type. This will lead to quantization of the surface anomalous Hall response superposed on conventional \textit{n}-type conduction.
	
	{Figures 2c-g show magnetotransport data for a sample (sample S1) irradiated with a $3.1~\textrm{C/cm}^2$ electron dose, and annealed at a series of temperatures $T_a$.
		The Hall conductance $G_{yx} = G_{yx}^N + G_{yx}^{AHE}$ (Fig.~2c) reflects the sum of the normal bulk response $G_{yx}^N$, which is linear in applied field, and the anomalous surface response $G_{yx}^{AHE}$, which shows a hysteretic step response around zero field.
		Consistent with the vacancy diffusion picture, the initial linear slope is large and negative ($n$-type). Upon annealing, it first decreases with increasing $T_a$, remains roughly constant when $T_a$ is in the $70-90^\circ \textrm{C}$ range, and then decreases again for higher $T_a$, finally converting to a positive ($p$-type) slope (also see {Fig.~S5)}.
		The overall downward trend with an apparent plateau  at intermediate annealing temperatures is seen clearly by plotting the Hall conductance at -1.5 T as a function of $T_a$ (Fig. 2d).
		The onset of the plateau in $G_{yx}$ corresponds to the point at which the Fermi level enters the 2DEG separation gap (point 2 in Figs. 2b and 2d).
		This is confirmed by measuring the gating response (Fig. 2e). The ambipolar behavior centered near zero voltage is direct evidence for type conversion \cite{ambipolar1-2014,ambipolar2-2016}, while the minimum in $R_{xx}$ points to the presence of chiral edge channels on the gapped surfaces \cite{QAH-realMat-2016,Yu-QAHtheory2010} coexisting with 2DEG longitudinal conductance channels.}
	
	Figure 2f shows the anomalous Hall conductance $G_{yx}^{AHE}$, obtained by subtracting the linear normal Hall conductance $G_{yx}^N$ from $G_{yx}$.  With increasing $T_a$, the height of the $G_{yx}^{AHE}$ loop first decreases, reaches a minimum half-height quantized value $G_0 = 1.001 e^2/h$ and then increases again.
	{Notably, the quantized value of $G_{yx}^{AHE}= G_0$, when $E_F$ and $\Delta$ align, coincides with the edge of the plateau in $G_{yx}$ (point 3 in Figs. 2b and 2d).
		Plotting $G_{yx}^{AHE}$ and the longitudinal conductivity $G_{xx}$ vs. $T_a$ (Fig. 2g) shows that the minimum in $G_{xx}$ slightly slightly precedes quantization of $G_{yx}$. This indicates that the minimum in the 2DEG density of states (DOS) is slightly above the Dirac gap (Fig. 2h). This behavior is seen in many samples, see e.g. Fig. S8.}}


Remarkably, contrary to previously reported QAH regime where $G_{xx}\rightarrow0$, {in our system at quantization $G_{xx}=\eta G_0$ is unexpectedly high ($\eta\sim 1000$).}
Here we illustrate with a simple summation of the 2D surface and bulk conductance channels how they are separable in 2D, which makes possible to detect quantization in the surface Hall (transverse) channel by tuning the Fermi level into the surface gap, even in the presence of a finite longitudinal conductance channel (see Supplementary Sections \textbf{C} and \textbf{D}).

The total conductance is a sum of the surface and bulk contributions, i.e. $G^{tot}_Q = G^{SS}_Q + G^{B}_Q$.
At quantization, when $E_F$ aligns with the Dirac mass gap, the conductance matrices are given by
$G^{SS}_Q = \frac{e^2}{h}
\begin{pmatrix}
0 & -1 \\
1 & 0
\end{pmatrix} ~ \textrm{and}~
G^{B}_Q = \frac{e^2}{h}
\begin{pmatrix}
\eta & 0 \\
0 & \eta
\end{pmatrix},$
which satisfy $G^{SS}_{xx}\rightarrow0$ {(dissipation-free  chiral edge conductance)} and $G^{SS}_{yx}\rightarrow G_0$ {({\emph{quantized}} Hall conductance, QAH)}
and enforce the experimental result that in the quantized regime there is minimal contribution to $G_{yx}$ from the bulk (see Figs.~3b,c below). 
The total resistance becomes
$R^{tot}_Q \cong \frac{h}{e^2}
\begin{pmatrix}
\frac{1}{\eta} & -\frac{1}{\eta^2} \\\textsl{}
\frac{1}{\eta^2} & \frac{1}{\eta}
\end{pmatrix}$,
{which  for $1/\eta \sim 10^{-3}$ is small, consistent with our experiments.
	Next we will show that, indeed, there exists a quantization window (`Q-window') in which the bulk contribution to $G_{yx}$ is zero.}

{The ease of tunability of the Fermi level and Dirac mass gap rests on the delicate balance between thickness and irradiation dose. The $T_a$ range, $\Delta T_a$,  needed to reach quantization --- in the putative Q-window --- is larger in thicker samples with lower or no electron irradiation.
	This is shown for three different samples in Fig.~3a.
	The bulk contribution to AHE is deduced by subtracting a quantized surface state (SS) contribution $G^{SS}_{yx}$ {(see Supplementary Sections \textbf{C} and \textbf{D}, also Ref. \cite{surface-states-SL-2019})} from the $G^{AHE}_{yx}$ data (e.g. for sample S1 in Fig.~3b),}
confirming the existence of the Q-window in the samples shown.
{A consistent presence of the Q-window is further seen in the samples that can not be tuned to quantization. This is illustrated for {(a thick, low-irradiation-dose)} sample S4, where within our window $G_{yx}$ is always less that $G_0$ (Figs.~3e,f). For this sample the surface bands are initially already in the BCB. With thermal tuning,
	{the vacancy distribution upshifts the gapped surface Dirac cone even deeper into the BCB, {so that only the tail} of $G_{yx}^{SS}$ is left within the regime where $G_{yx}^B\sim 0$. When the Fermi level resides in the $G_{yx}^B \sim 0$ window, then $G_{yx}$ is much much smaller than $G_0$, as neither the bulk nor surface contribute significantly (Figs.~3c,d). }
	{In this sample, the Hall slope switches from negative to positive (\textit{n-}to-\textit{p} conduction type transition) at low magnetic fields while remaining negative at high fields (Figs.~3e). This is clearly not an ambipolar conversion across CNP --- the high-field negative slope of $G_{yx}$ indicates that bulk is still \textit{n}-type when the conversion is observed  --- it is when {\textit{p}-type surface carriers compensate \textit{n}-type bulk (Fig.~3d).}
		Here, $E_F$  is still within BCB, but $G_{yx}$ becomes tiny ($\leq 0.04~ G_{0}$), confirming that the bulk contribution within the window is indeed negligible.}}


{To further test this regime we add the bulk contribution deduced from sample S1 to that from SS to obtain the total $G_{xy}^{AHE}$ for when SS is at
	the edge of the Q-window and $E_F$ still aligned with $\Delta$. This is the case of the unirradiated sample S3, where {$G_{yx}$ is large (several $G_0$) to start and both extrinsic and intrinsic contributions \cite{AHE-Nagaosa2010} to $G_{yx}$ are possible. Affirming Q-window, $G_{xy}^{AHE}$ quickly drops near quantization (Q), eventually reaching the value of $G_{0}$.
		Indeed, the total {deduced} $G_{yx}^{AHE}$ closely mimics the observed bulk features once the energy range is rescaled by $\Delta T_a$ (see Fig.~4a).}
	Here, once the surface bands are relocated to the vicinity of the Q-window, we can explore and fine-tune $G_{yx}$ by electrostatic gating. $G_{yx}$ vs. gate  voltage $V_g$ at zero magnetic field displays a true $e^2/h$ plateau, which is completely absent when the gating range is shifted beyond the Q-window (Fig.~4b). In a finite field,  $G_{yx}(V_g)$ exhibits a quasi-plateau (Fig.~4c) --- similar to that observed under thermal tuning --- {reflecting the field dependence of the surface gap (Figs.~4d,e and S11)).}}

{The fundamental reason for the existence of the Q-window is that Berry curvature of the bulk bands is concentrated around the small gaps (`avoided crossings') opened by spin-orbit coupling \cite{BerryPhaseReview2010}, quickly becoming zero away from them. {Thus, when the surface Dirac cone resides within the bulk conduction band, under tunable bandstructure conditions, the Hall conductance $\sigma_{yx}$ will settle on a plateau to a great precision around its quantized value.} {{As a proof-of-principle, a simple model calculation of
			$\sigma_{yx}^{AHE} = \frac{e^2}{\hbar} \int_{BZ} \frac{d^2 k}{(2\pi)^2} \sum_n \Omega^{nk}_{k_xk_y}f(\epsilon_{nk} - \epsilon_F)$ captures the key observed behaviors, namely the presence of the $e^2/h$ plateau as well as the anomalous Hall conductance trend outside the Q-window (inset in Fig. 4b). 
			Crucially, this happens without fine-tuning when there is an energy window where the 2D bulk $G_{yx} = 0$, which allows $G_{yx}$ to attain the quantized surface value of $e^2/h$ {(Figs. S10-S12)}.}} In our calculation $\sigma_{yx}^{AHE}$ is obtained by integrating the Berry curvature $\Omega^{nk}$ with the Fermi-Dirac distribution $f(\epsilon_{nk} - \epsilon_F)$ 
	over the Brillouin zone (BZ) of the $n$ bands corresponding to our superlattice (Supplementary Section \textbf{C} and {Figs. S9, S10}).
	
	The Q-window uncovered here is {eminently} tunable --- it can be expanded and further controlled on-demand by optimizing the initial (nonmagnetic) TI bandstructure so that the Dirac point is well within the bulk gap. The intrinsic superlattice magnetic TIs can be grown with different magnetic elements \cite{surface-states-SL-2019} within the superlattice sequences to increase $T_C$, potentially to be much above the temperature range demonstrated here.}
Our new tuning toolbox --- electron irradiation {and thermal vacancy redistribution} ---  lifts the thickness restriction in a wide library of this new materials class.
The uncovered QAH regime is robust and can be distinguished even in the presence of bulk carriers for materials over an order of magnitude thicker than the $\lesssim~10$ nm upper limit considered thus far. With lower magnetic doping disorder it could be advantageous for novel device structures,
and as a platform for building QAH-superconductor systems.

\vspace{3mm}
\noindent {\textbf{Methods}\\
	\def\baselinestretch{1.2}
	\small \noindent {\underline{Crystal growth and structural characterization}.}
	Crystals of Bi$_2$Te$_3$ alloyed with Mn were grown by the vertical Bridgman method following the two-step technique. 
	In the first step the material was synthesized. Ground (20-100~mesh) high purity components (99.999\%), bismuth (Bi), tellurium (Te) and manganese (Mn), were weighted according to the formula Mn$_x$Bi$_{2-x}$Te$_3$ and loaded into the double quartz ampules to avoid depressurization during the cooling process. The total weight of the charge per ampule was 35-40~g. The ampules were evacuated to $10^{-6}$ Torr and sealed. The prepared ampules were loaded into a vertical furnace, heated to 900~K, and maintained at this temperature for 48~h to achieve better homogenization. After, it was cooled down to 550~K  (speed 60~K per hour) and annealed for 24 hours. Then the furnace was switched off and cooled down to room temperature.
	In the second step, the synthesized single phase material was ground again and loaded into the ampules for the Bridgman growth, evacuated to $10^{-6}$ Torr and sealed. The special feature of the growth ampules is a small diameter (1.5-2.0~mm) along the tip in the lower end of the ampules used to create the seed crystal. Two ingots obtained from the first (synthesis) step were used to fill the Bridgman growth ampule. To obtain a homogenized solution, the growing material was heated to 1073 K and rotated along the ampule axis for five days in the hot part  of the furnace. Then samples were moved down from the hot part of the furnace at the speed of 2~mm per day. The temperature in the lower part of the furnace was kept at 873 K. {This procedure resulted in \textit{n}-type crystals of Mn$_x$Bi$_{2-x}$Te$_3$ with average sizes of 50~mm length and 14~mm diameter. It should be noted that the single step process \cite{Hor-ptype_MnBT-2010} resulted in the \textit{p}-type crystals (see {Figs.~S4, S5}). }
	
	\noindent \underline{Electron irradiation}. Electron irradiations were carried out in a NEC Pelletron-type electrostatic accelerator at the Laboratoire de Physique des Solides at \'{E}cole Polytechnique, Palaiseau, configured with a low-temperature target maintained at 20 K in a chamber filled with liquid hydrogen fed from a close-cycle refrigerator \cite{irrad-Lukas2016}.
	All irradiations were performed with samples kept at 20 K, below the mobility threshold of the interstitials which tend to be more mobile than vacancies
	-- this ensured the stability of all charges introduced by the irradiation process.
	The beam current density, typically 2~$\mu$A on a $0.2~\textrm{cm}^2$,
	allowed modifications of carrier concentration on the order of
	$10^{20}~\textrm{cm}^{-3}$.
	
	\noindent\underline{Magnetic and transport measurements}.
	dc Magnetization measurements {of the single crystal samples were performed in the Superconducting Quantum Interference Device (SQUID) magnetic property measurement system (MPMS).}
	{Transport measurements were performed in a 14 Tesla Quantum Design Physical property measurement system (PPMS) in 1 Torr (at low temperature) of He gas on many samples, each subjected to the same annealing protocol. The samples were annealed \textit{in situ} starting from 330 K to 400 K.
		The annealing temperature ramp rate was 7 K per minute with the annealing time typically $\sim1$ hour. For $T_{a} > 400 K$, samples were annealed \textit{ex situ} in a vacuum furnace.
		Crystals were mechanically exfoliated onto 300 nm SiO$_2$/Si$^{+++}$ wafers, typically resulting in micron-size crystals with thicknesses less than $\sim 400~\textrm{nm}$, as determined by the atomic force microscope (AFM).  Electrical contacts in the van der Pauw (vdP) configuration
		were photo-lithographically patterned and a sputtered Au metallurgy was used (Fig.~2d).
		The VdP dc measurements were carried out on hundreds of samples using a custom-configured system;
		reversing current direction was employed for each measurement to minimize thermal \textit{emf}. For {further} sample characterization see Supplementary Section \textbf{A}.
		\small
		\vspace{2mm}
		\noindent \textbf{Acknowledgements} We wish to acknowledge {Andy Millis} for his helpful insights and {Jim Hone} for the critical reading of the manuscript. This work was supported by the NSF grants DMR-1420634 (Columbia-CCNY MRSEC) and HRD-1547830, {and by the National Science Center (Poland), grant 2016/21/B/ST3/02565.}
		Computational support was provided by Virginia Tech ARC and San Diego Supercomputer Center (SDSC) under DMR-060009N.
		
		\vspace{2mm}
		\noindent\textbf{Author Contributions} Experiments were designed by L.K.-E. and H.D.. Device fabrication, transport, and magnetic measurements were performed by H.D.. Structural (TEM) and elemental characterization of the crystals grown by I.V.F. was done by K.S. and J.S.. A.W. characterized samples by ferromagnetic resonance (FMR). Electron irradiations at LSI were conducted by M.K. with the assistance of Z.C. and H.D. ARPES studies were performed by T. H. and L.P. DFT bandstructure was calculated by K.P.. A.B.G. and J.C. calculated AHE conductance from the Berry curvature.  Data analysis was done by H.D. and L.K.-E. L.K.-E. wrote the manuscript with critical input from H.D.
		
		
\def\baselinestretch{1.28}
\vspace{-6mm}
\bibliographystyle{naturemag}
\bibliography{refsMain173,booksetc5}

\newpage
\noindent\section*{FIGURE LEGENDS}
\vspace{-3mm}
\normalsize
\noindent \textbf{Figure 1 $\mid$ Quantized Hall conductance in a bulk magnetic superlattice structure of a dilute Mn-doped Bi$_2$Te$_3$.}
\textbf{a}, Periodic sequence of Bi$_2$Te$_3$ (QLs) and MnBi$_2$Te$_4$ (SLs).
\textbf{b}, Zoom of (\textbf{a}). Mn monolayers within SLs imaged by high-resolution scanning TEM (\textit{left}) and EDX (\textit{right}), see Supplementary file.
\textbf{c}, Crystal structure of SLs in the alternating QL-SL sequence.
\textbf{d}, {Illustration of the out-of-plane magnetization $M$, with 1D chiral channels on the edges of the 2D surfaces of a 3D TI.
	\textbf{e}, $M(T)$ (red) measured in {a -100 Oe field}. $G_{yx}^{AHE}(T)$ (blue) onsets at $T_C \cong 13~\textrm{K}$, and begins to approach the quantized value $e^2/h$ from about 7~K on cooling.}
\textbf{f}, $G_{yx}^{AHE}(H)$ hysteresis loop  at 1.9~K  {with a coercive field $H_c \approx 50~\textrm{mT}$}  and the zero-field 1/2 the loop height $G_0 = e^2/h$.
\textbf{g}, The corresponding longitudinal conductance $G_{xx}^{AHE}(H)$.
For reciprocal conductance-to-resistance tensor conversion see Supplementary Section \textbf{C}.

\vspace{3mm}
\noindent \textbf{Figure 2 $\mid$ Tuning the bulk magnetic superlattice structure to quantization: irradiation plus thermal annealing}.
{\textbf{a}, Vacancy redistribution by thermal annealing establishes vacancy density gradient, with surface regions (blue shade) progressively denuded of vacancies.
	\textbf{b}, The reduced vacancy density in the bulk downshifts the initial Fermi energy $E_F^0$ from BCB (left, arrow down) and the band bending of the subsurface regions upshifts the Dirac and 2DEG bands relative to BCB (right, arrow up), as shown in four representative {steps}. Quantization (Q) will realize when three relevant energy scales --- the Fermi level, the Dirac gap $\Delta$,  and the  separation of 2DEG bands $E_{2D}$ --- will align. $E_F^Q$ is the Fermi level at quantization. Bulk valence bands are marked as BVB. \textbf{c}, $G_{yx}(H)$ for different annealing temperatures $T_a$. Initially, the normal Hall conductance $G_{yx}^N(H)$ is $n$-type and nearly field-linear in the range shown. It monotonically decreases with $T_a$  until it hits a snag on an apparent plateau in
	(\textbf{d}), prior to the eventual conversion to $p$-type (negative $G_{yx}$).
	\textit{Inset}: {Optical image of van der Pauw contact geometry used.}
	\textbf{e}, Voltage gating at Q reveals ambipolar behavior in Hall resistance $R_{yx}$ riding on the bulk background. $R_{xx}$ shows a minimum expected in a chiral state.
	\textbf{f}, $G_{yx}^{AHE}(H)$ hysteresis loops at different $T_a$. Quantization is reached at $T_a = 92^\circ\textrm{C}$. \textbf{g}, $G_{yx}^{AHE}$ and $G_{xx}$ as a function of $T_a$. {A dip in $G_{xx}(T_a)$ corresponds to a reduced density of states (DOS) {within $E_{2D}$ (blue shade}), see (\textbf{h}}). The dip is slightly shifted from the $G_{yx}^{AHE} \cong G_{0}$ owing to a misalignment of DOS minimum and the Dirac gap $\Delta$.
	The schematic annealing steps in (\textbf{d}) and (\textbf{g}) are numbered as in (\textbf{b}).}} {\textbf{h}, DOS of the \textit{n}-type bulk bands (blue) and surface \textit{n}-type (green) and \textit{p}-type (red) bands at Q. The minimum in the bulk DOS is within $E_{2D}$.}

\vspace{3mm}
\noindent \textbf{Figure 3 $\mid$ Quantization window for a finite \emph{n}-type $G_{xx}$.}
\textbf{a}, $G_{yx}^{AHE}$ \textit{vs}. the annealing temperature interval ($\Delta T_a$) needed to arrive at quantization (Q), normalized to $T_a$ at Q and to samples' thickness $t$ for samples S1 {($t = 300~\textrm{nm}$)} and S2 {($t = 70~\textrm{nm}$)} with irradiation doses of 3.1 and $0.75 ~\textrm{C/cm}^2$ respectively, and for the unirradiated sample S3 {($t = 101~\textrm{nm}$)}, where a 		{larger initial $G_{yx}$ derives from the bulk}. 
\textit{Inset}: Bandstructure cartoon at Q.
{\textbf{b}, $G_{yx}^{AHE} = G_0$ at Q has a minimum since the contribution to Hall conductance from the (3D) bulk (grey shade) drops precipitously near Q and the contribution from the subsurface 2DEG (purple shade) is shown to vanish within a Q-window within BCB, see text.  Both contributions add to the intrinsic Berry contribution from the surface states (SS), which peaks at $G_0$.
The quantization is observed within this Q-window when the surface gap $\Delta$ is aligned with $E_F$ (vertical dash), here illustrated for sample S1.
\textbf{c}, When the initial location of $\Delta$ is deep in the BCB above the Q-window  but $E_F$ is still within it, only a tail ($\ll G_0$) of SS contributes to  $G_{yx}^{AHE}$.
Red arrows indicate the direction of the SS upshift (from red dash to solid).
{\textbf{d}, DOS sketch illustrates how \emph{n}-type bulk carriers ($G_{xx}  \neq 0$) can be compensated by the \emph{p}-type surface carriers.}
\textbf{e}, {$G_{yx}(H)$ of a non-quantized sample S4 ($t = 300~\textrm{nm}$, $\phi \cong 0.5~\textrm{C/cm}^2$ ).} The \textit{n}-to-\textit{p} transition  (negative-to-positive $G_{yx}(H)$ slope) is seen only at low magnetic fields, indicating that the bulk is still \textit{n}-type.
\textbf{f},} The surface \textit{n}-to-\textit{p} conversion {in sample S4}, where the Q-window and the surface $\Delta$ do not align and the quantization is not observed. It demonstrates that in the Q-window the contribution from the bulk is indeed very {small ($\leq 0.04~ G_{0}$)}.
Labels \textbf{1} and \textbf{2} correspond to $T_a = 48^\circ \textrm{C ~and} ~ 127^\circ \textrm{C}$ respectively.

\vspace{3mm}
\noindent \textbf{Figure 4 $\mid$ Gating dependence of Berry-curvature-driven QAH.}
\textbf{a}, The bulk contribution to $G_{yx}^{AHE}$ from Fig.~3B added to the SS contribution (red) for SS upshifted to the edge of BCB but still aligned with $E_F$. The structure in the total $G^{AHE}_{yx}$ arising from the bulk mimics that seen in sample S3 (blue squares) once the temperature range is rescaled to $\Delta T_a$.
\textbf{b}, Quantization plateau in $G_{yx} (V_g)$ at zero field when $E_F$ {in sample S3} is aligned with $\Delta$ (red circles).  The plateau is not observed when $E_F$ is outside the Q-window (grey circles). {\textit{Inset}: Model calculation under finite surface voltage (Supplementary Section \textbf{D}) confirms the existence of the $e^2/h$ plateau and reproduces the behavior of $G_{yx}$ outside the Q-window.
\textbf{c}, $G_{yx}(V_g)$ exhibits {a quasi-plateau akin to that obtained by annealing}.
\textit{Lower inset}: Cartoon of the surface and bulk states on the quasi-plateau.
{\textit{Upper inset}: 3D contour plot of $\frac{\textrm{d}G_{yx}(V_g, H)}{\textrm{d}V_g }$  --- a measure of the Dirac gap $\Delta$ which widens with applied magnetic field (see also {Fig.~S13}).}
\textbf{d}, Subtracting the high-field bulk contribution $G^B_{yx}$ from the total $G_{yx}$, gives $\Delta G_{yx}(V_g)\cong~G_0 = {e^2}/{h}$.

\newpage

\hspace{-14mm}
\includegraphics[width=1.08\textwidth]{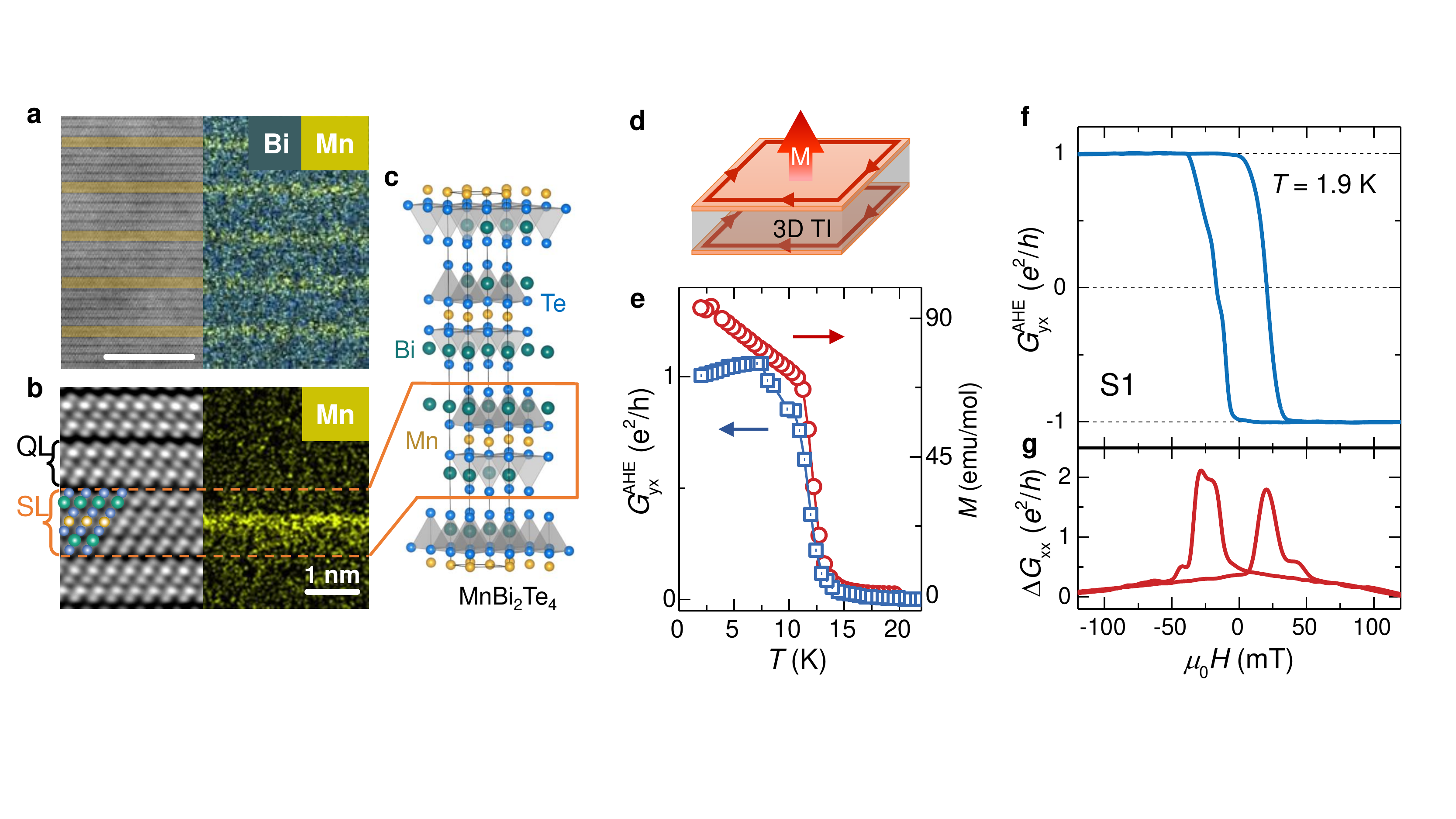}
\vfill\hfill Fig.~1; {HD \it et al.} \eject

\includegraphics[width=13.9cm]{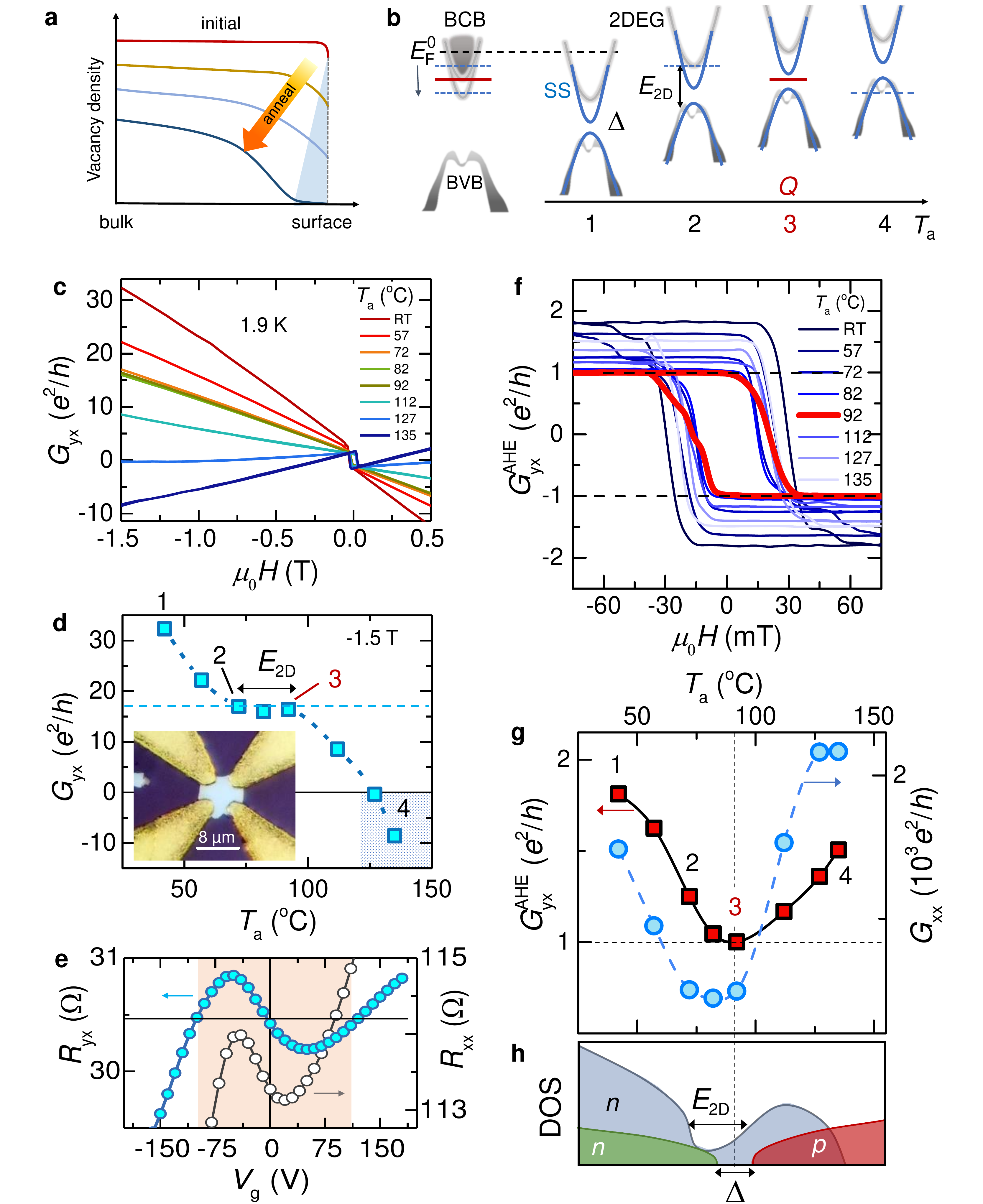}
\vfill\hfill Fig.~2; {HD \it et al.} \eject

\hspace{-18mm}
\includegraphics[width=1.09\textwidth]{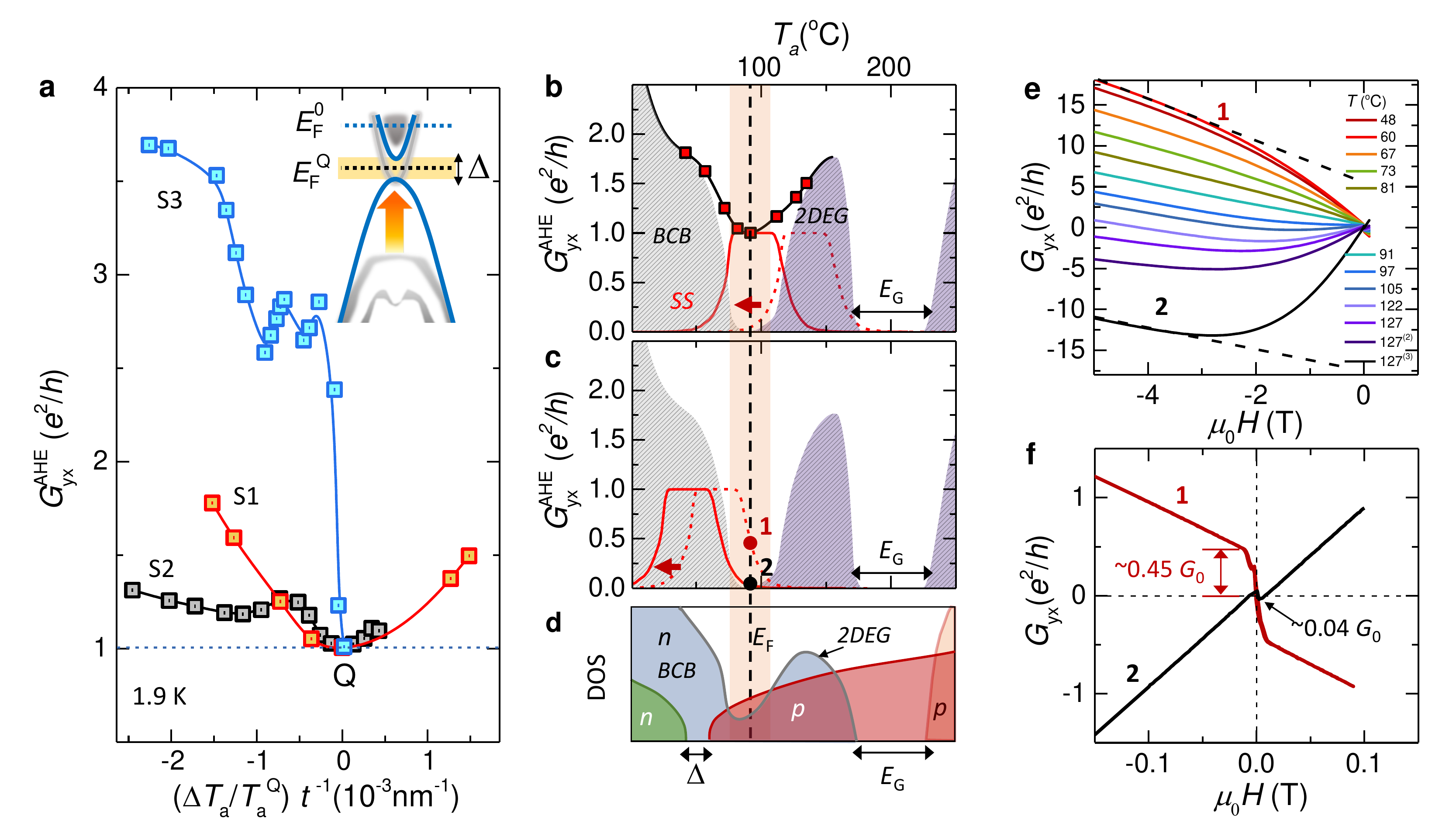}
\vfill\hfill Fig.~3; {HD \it et al.} \eject

\includegraphics[width=0.9\textwidth]{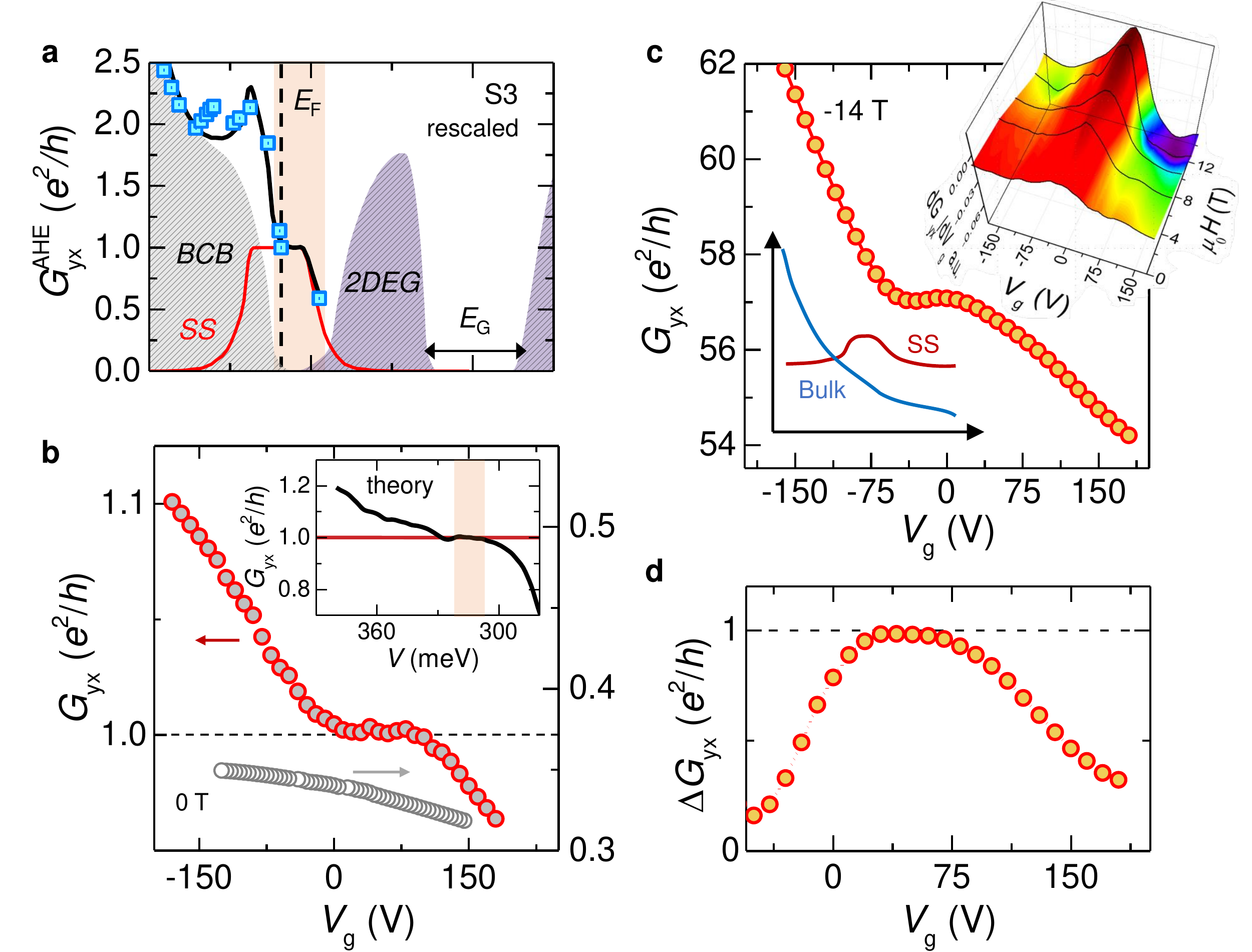}
\vfill\hfill Fig.~4; {HD \it et al.} \eject

\end{document}